\documentclass[twocolumn,showpacs,preprintnumbers,amsmath,amssymb]{revtex4}
\usepackage{graphicx}% Include figure files
\usepackage{dcolumn}% Align table columns on decimal point
\usepackage{bm}% bold math
\usepackage{epsfig}

\begin{document}

\preprint{DESY~09--144\hspace{14.2cm} ISSN 0418--9833}
\preprint{September 2009\hspace{16.3cm}}

\title{\boldmath Complete next-to-leading-order corrections to $J/\psi$
photoproduction in nonrelativistic quantum chromodynamics}

\author{Mathias Butensch\"on}
%\email{mathias.butenschoen@desy.de}
\author{Bernd A. Kniehl}
%\email{bernd.kniehl@desy.de}
\affiliation{{II.} Institut f\"ur Theoretische Physik, Universit\"at Hamburg,
Luruper Chaussee 149, 22761 Hamburg, Germany}

\date{\today}

\begin{abstract}
We calculate the cross section of inclusive direct $J/\psi$ photoproduction at
next-to-leading order within the factorization formalism of nonrelativistic
quantum chromodynamics, for the first time including the full relativistic
corrections due to the intermediate $^1\!S_0^{[8]}$, $^3\!S_1^{[8]}$, and
$^3\!P_J^{[8]}$ color-octet states.
A comparison of our results to recent H1 data suggests that the color octet
mechanism is indeed realized in $J/\psi$ photoproduction, although the
predictivity of our results still suffers from uncertainties in the
color-octet long-distance matrix elements.
\end{abstract}

\maketitle

%Since the discovery of the $J/\psi$ meson in 1974, charmonium has provided a
%useful laboratory for quantitative tests of quantum chromodynamics (QCD) and,
%in particular, of the interplay of perturbative and nonperturbative phenomena.
The factorization formalism of nonrelativistic %QCD
quantum chromodynamics (NRQCD) \cite{Bodwin:1994jh}
provides a consistent theoretical framework for the description of
heavy-quarkonium production and decay, which is known to hold through two loops
\cite{Nayak:2005rt}.
This implies a separation of process-dependent short-distance coefficients, to
be calculated perturbatively as expansions in the strong-coupling constant
$\alpha_s$, from supposedly universal long-distance matrix elements
(LDMEs), to be extracted from experiment.
The relative importance of the latter can be estimated by means of velocity
scaling rules; {\it i.e.}, the LDMEs are predicted to scale with a definite
power of the heavy-quark ($Q$) velocity $v$ in the limit $v\ll1$.
In this way, the theoretical predictions are organized as double expansions in
$\alpha_s$ and $v$.
A crucial feature of this formalism is that it takes into account the complete
structure of the $Q\overline{Q}$ Fock space, which is spanned by the states
$n={}^{2S+1}L_J^{[a]}$ with definite spin $S$, orbital angular momentum
$L$, total angular momentum $J$, and color multiplicity $a=1,8$.
In particular, this formalism predicts the existence of color-octet (CO)
processes in nature.
This means that $Q\overline{Q}$ pairs are produced at short distances in
CO states and subsequently evolve into physical, color-singlet (CS) quarkonia
by the nonperturbative emission of soft gluons.
In the limit $v\to0$, the traditional CS model (CSM) is recovered in the case
of $S$-wave quarkonia.
In the case of $J/\psi$ production, the CSM prediction is based just on the
$^3\!S_1^{[1]}$ CS state, while the leading relativistic corrections, of
relative order ${\cal O}(v^4)$, are built up by the $^1\!S_0^{[8]}$,
$^3\!S_1^{[8]}$, and $^3\!P_J^{[8]}$ ($J=0,1,2$) CO states.

Fifteen years after the introduction of the NRQCD factorization formalism
\cite{Bodwin:1994jh}, the existence of CO processes and the universality of the
LDMEs are still at issue and far from proven, despite an impressive series of
experimental and theoretical endeavors.
The greatest success of NRQCD was that it was able to explain the $J/\psi$
hadroproduction yield at the Fermilab Tevatron \cite{Cho:1995vh}, while the
CSM prediction lies orders of magnitudes below the data, even if the latter
is evaluated at next-to-leading order (NLO) or beyond
\cite{Campbell:2007ws,Gong:2008sn}.
Also in the case of $J/\psi$ photoproduction at DESY HERA, the CSM cross
section significantly falls short of the data, as demonstrated by a recent NLO
analysis \cite{Artoisenet:2009xh} using up-to-date input parameters and
standard scale choices, leaving room for CO contributions
\cite{Cacciari:1996dg}.
Similarly, the $J/\psi$ yields measured in electroproduction at HERA and in
two-photon collisions at CERN LEP2 were shown
\cite{Kniehl:2001tk,Klasen:2001cu} to favor the presence of CO processes.
%On the other hand, the NLO CSM cross sections of $J/\psi+c\overline{c}$
%\cite{Zhang:2006ay} and $J/\psi+gg$ \cite{Ma:2008gq} describe data from the $B$
%factories much better than the leading-order (LO) predictions, leaving lesser
%room for CO contributions.
As for $J/\psi$ polarization in hadroproduction, neither the %LO
leading-order (LO) NRQCD
prediction \cite{Braaten:1999qk}, nor the NLO CSM one \cite{Gong:2008sn} leads
to an adequate description of the Tevaton data.
The situation is quite similar for the polarization in photoproduction at HERA
\cite{Artoisenet:2009xh}.

In order to convincingly establish the CO mechanism and the LDME universality,
it is an urgent task to complete the NLO description of $J/\psi$ hadro-
\cite{Campbell:2007ws,Gong:2008sn,Gong:2008ft} and photoproduction
\cite{Artoisenet:2009xh,Kramer:1994zi}, regarding both $J/\psi$ yield
\cite{Campbell:2007ws,Kramer:1994zi} and polarization
\cite{Gong:2008sn,Artoisenet:2009xh,Gong:2008ft}, by including the full CO
contributions at NLO.
While the NLO contributions due to the $^1\!S_0^{[8]}$ and $^3\!S_1^{[8]}$
CO states may be obtained \cite{Gong:2008ft} using standard techniques,
familiar from earliest NLO CSM calculations \cite{Kramer:1994zi}, the NLO
treatment of $^3\!P_J^{[8]}$ states in $2\to2$ processes requires a more
advanced technology, which has been lacking so far.
In fact, the $^3\!P_J^{[8]}$ contributions represent the missing links in all
those previous NLO analyses
\cite{Campbell:2007ws,Gong:2008sn,Artoisenet:2009xh,Gong:2008ft,Kramer:1994zi},
and there is no reason at all to expect them to be insignificant.
Specifically, their calculation is far more intricate because the application
of the $^3\!P_J^{[8]}$ projection operators to the short-distance scattering
amplitudes produce particularly lengthy expressions involving complicated
tensor loop integrals and exhibiting an entangled pattern of infrared (IR)
singularities.
This technical bottleneck, which has prevented essential progress in the global
test of NRQCD factorization for the past fifteen years, is overcome here for
the first time.
So far, only two complete NLO analyses of heavy-quarkonium production in
high-energy collisions involving CO states have been performed:
the total cross section of hadroproduction \cite{Petrelli:1997ge} and the
inclusive cross section at finite transverse momentum $p_T$ in two-photon
collisions \cite{Klasen:2004tz}.
However, the former case corresponds to a $2\to1$ process, which enormously
simplifies the calculation, and the latter case does not involve virtual
corrections in $P$-wave channels.

\begin{figure}
\includegraphics[width=5cm]{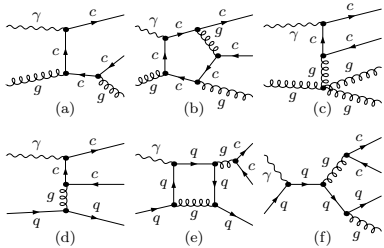}
\caption{\label{fig:Examples} Sample diagrams contributing at LO (a and d) and
to the virtual (b and e) and real (c and f) NLO corrections.}
\end{figure}

In direct photoproduction, a quasi-real photon $\gamma$ that is radiated off
the incoming electron $e$ interacts with a parton $i$ stemming from the
incoming proton $p$.
Invoking the Weizs\"acker-Williams approximation %\cite{von Weizsacker:1934sx}
and the factorization theorems of the QCD parton model and NRQCD
\cite{Bodwin:1994jh}, the inclusive $J/\psi$ photoproduction cross section is
evaluated from
\begin{eqnarray}
d\sigma(ep\to J/\psi+X)
&=&\sum_{i,n} \int dxdy\, f_{\gamma/e}(x)f_{i/p}(y)
\label{Overview.Cross}\\
&&{}\times\langle{\cal O}^{J/\psi}[n]\rangle
d\sigma(\gamma i\to c\overline{c}[n]+X),
\nonumber
\end{eqnarray}
where $f_{\gamma/e}(x)$ is the photon flux function, $f_{i/p}(y)$ are the
parton distribution functions (PDFs) of the proton,
$\langle{\cal O}^{J/\psi}[n]\rangle$ are the LDMEs, and
$d\sigma(\gamma i\to c\overline{c}[n]+X)$ are the partonic cross sections.
Working in the fixed-flavor-number scheme, $i$ runs over the gluon $g$ and the
light quarks $q=u,d,s$ and anti-quarks $\overline q$.
%The Fock states contributing through the order of our calculation include
%$n={}^3S_1^{[1]},{}^1S_0^{[8]},{}^3S_1^{[8]},{}^3P_J^{[8]}$.

The Feynman amplitudes of $\gamma i\to c\overline{c}[n]+X$ are calculated by
the application of appropriate spin and color projectors onto the usual Feynman
amplitudes for open $c\overline{c}$ production \cite{Petrelli:1997ge}.
Example Feynman diagrams for partonic LO subprocesses as well as virtual- and
real-correction diagrams are shown in Fig.~\ref{fig:Examples}.
Important properties of these projections are that the relative momentum $q$
between the $c$ and $\overline{c}$ quarks has to be set to zero, in the case
of $P$-wave states after taking the derivative with respect to $q$.

\begin{figure}
\includegraphics[width=8cm]{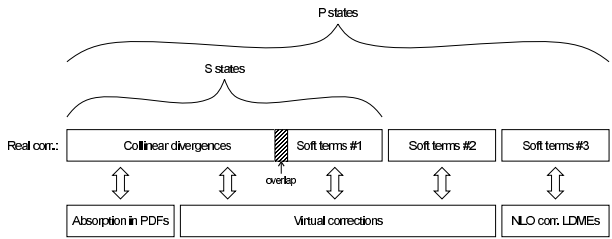}
\caption{\label{SingStruct}Overview of the IR singularity structure.}
\end{figure}

We checked analytically that all appearing singularities cancel.
As for the ultraviolet singularities, we renormalize the charm-quark mass and
the wave functions of the external particles according to the on-shell scheme,
and the strong-coupling constant according to the modified minimal-subtraction
scheme.
Figure~\ref{SingStruct} displays an overview of the IR singularity
structure.
In the case of the ${}^3S_1^{[1]}$, ${}^1S_0^{[8]}$, and ${}^3S_1^{[8]}$
states, the soft and collinear singularities of the real corrections are
canceled as usual by complementary contributions stemming from the virtual
corrections and by the absorption of universal parts into the proton and photon
PDFs, the latter entering via resolved photoproduction.
In case of the $^3\!P_J^{[8]}$ states, the soft singularity structure is more
complex.
The reason is the following:
In the soft limit, the real-correction amplitudes factorize into LO amplitudes
and so-called eikonal factors.
Taking the derivative with respect to $q$ and squaring the amplitudes then
leads to additional {\em soft~\#2} and {\em soft~\#3} terms because the
derivative has to be taken of the eikonal factors as well.
The soft~\#3 terms are proportional to a linear combination of the
short-distance cross sections to produce the ${}^3\!S_1^{[1]}$ and
${}^3\!S_1^{[8]}$ states.
They are canceled against IR singularities stemming from radiative corrections
to the $\langle {\cal O}^{J/\psi}(^3\!S_1^{[1]}) \rangle$ and
$\langle {\cal O}^{J/\psi}(^3\!S_1^{[8]}) \rangle$ LDMEs.
The soft~\#2 terms do not factorize to LO cross sections.
They also cancel against virtual-correction contributions as the usual soft~\#1
terms.

Apart from the analytical cancellation of all occurring singularities, our
calculation passes a number of further nontrivial checks.
We implemented two independent methods for the reduction of the tensor loop
integrals, which yielded identical results.
As for the real corrections, the numerical evaluation of our expressions for
the squared matrix elements agree with numerical output generated using the
program package MadOnia \cite{Artoisenet:2007qm},
%, a quarkonium version of the program package MadGraph \cite{Maltoni:2002qb}, 
well within the numerical uncertainty of the latter.
We verified that our results are stable w.r.t.\ variations of the phase space
slicing parameters introduced as a demarcation between the soft and/or
collinear regions from the rest of the three-particle phase space.
\cite{Harris:2001sx}.
Finally, we could nicely reproduce the NLO CSM results of
Ref.~\cite{Kramer:1994zi} after adopting the inputs chosen therein.
For space limitation, we refrain from presenting here more technical details,
but refer the interested reader to a forthcoming publication.

\begin{figure}
\includegraphics[width=5cm]{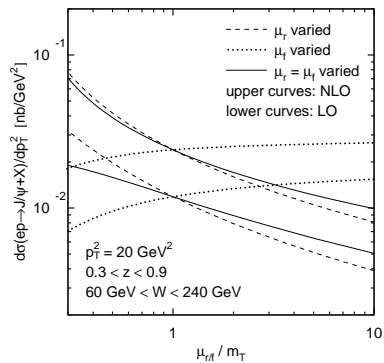}
\caption{\label{fig:scales}Separate and joint dependences of
$d\sigma(ep\to J/\psi+X)/dp_T^2$ at $p_T^2=20$~GeV$^2$ in full NRQCD at LO and
NLO on $\mu_r$ and $\mu_f$.}
\end{figure}

We now describe our theoretical input and the kinematic conditions for our
numerical analysis.
We set $m_c=m_{J/\psi}/2$, adopt the values of $m_{J/\psi}$, $m_e$, and
$\alpha$ from Ref.~\cite{Amsler:2008zzb}, and use the one-loop (two-loop)
formula for $\alpha_s^{(n_f)}(\mu)$, with $n_f=3$ active quark flavors, at LO
(NLO).
As for the proton PDFs, we use set CTEQ6L1 (CTEQ6M) \cite{Pumplin:2002vw} at LO
(NLO), which comes with an asymptotic scale parameter of
$\Lambda_\mathrm{QCD}^{(4)}=215$~MeV (326~MeV), so that
$\Lambda_\mathrm{QCD}^{(3)}=249$~MeV (389~MeV).
We evaluate the photon flux function using Eq.~(5) of Ref.~\cite{Kniehl:1996we}
with the cut-off $Q_\mathrm{max}^2=2$~GeV$^2$ \cite{Adloff:2002ex,H1.prelim} on
the photon virtuality.
Our default choices for the renormalization, factorization, and NRQCD scales
are $\mu_r=\mu_f=m_T$ and $\mu_\Lambda=m_c$, respectively, where
$m_T=\sqrt{p_T^2+4m_c^2}$ is the $J/\psi$ transverse mass.
We adopt the LDMEs from Ref.~\cite{Kniehl:1998qy}, which were fitted to
Tevatron~I data using the CTEQ4 PDFs, because, besides the usual LO set, they
also comprise a {\em higher-order-improved} set determined by approximately
taking into account dominant higher-order effects due to multiple-gluon
radiation in inclusive $J/\psi$ hadroproduction, which had been found to be
substantial by a Monte Carlo study \cite{CanoColoma:1997rn}.
This observation is in line with the sizable NLO corrections recently found
in Refs.~\cite{Campbell:2007ws,Gong:2008sn,Gong:2008ft}, still excluding the
$^3\!P_J^{[8]}$ channels at NLO.
Of course, LDME fits to more recent Tevatron data are available, but their
goodness is clearly limited by the present theoretical uncertainties in the
short-distance cross sections, preventing the increase in experimental
precision gained since the analysis of Ref.~\cite{Kniehl:1998qy} from
actually being beneficial.
Apart from that, the central values of the $J/\psi$ LDMEs have only moderately
changed, as may be seen by comparing the LO results of
Ref.~\cite{Kniehl:1998qy} with those recently obtained \cite{Kniehl:2006qq} by
fitting Tevatron~II data using the CTEQ6L1 PDFs \cite{Pumplin:2002vw}.
Because the $p_T$ distributions of the $^1\!S_0^{[8]}$ and $^3\!P_J^{[8]}$
contributions to $J/\psi$ hadroproduction exhibit very similar shapes, fits
usually only constrain the linear combination
\begin{equation}
M_r^{J/\psi} = \langle {\cal O}^{J/\psi}(^1\!S_0^{[8]}) \rangle 
+ \frac{r}{m_c^2} \langle {\cal O}^{J/\psi}(^3\!P_0^{[8]}) \rangle,
\label{eq:mr}
\end{equation}
with an $r$ value of about 3.5 \cite{Kniehl:1998qy,Kniehl:2006qq}.
As in Ref.~\cite{Klasen:2004tz}, we take the democratic choice
$\langle {\cal O}^{J/\psi}(^1\!S_0^{[8]}) \rangle 
= (r/m_c^2) \langle {\cal O}^{J/\psi}(^3\!P_0^{[8]}) \rangle
=M_r^{J/\psi}/2$ as our default.

Recently, the H1 Collaboration presented preliminary data on inclusive $J/\psi$
photoproduction taken in collisions of 27.6~GeV electrons or positrons on
920~GeV protons in the HERA~II laboratory frame \cite{H1.prelim}.
They nicely agree with their previous measurement at HERA~I
\cite{Adloff:2002ex}.
These data come as singly differential cross sections in $p_T^2$,  
$W=\sqrt{(p_\gamma+p_p)^2}$, and $z=(p_{J/\psi}\cdot p_p)/(p_\gamma\cdot p_p)$,
in each case with certain acceptance cuts on the other two variables.
Here, $p_\gamma$, $p_p$, and $p_{J/\psi}$ are the photon, proton, and $J/\psi$
four-momenta, respectively.
In the comparisons below, we impose the same kinematic conditions on our
theoretical predictions.

\begin{figure*}
\begin{tabular}{ccc}
\includegraphics[width=5cm]{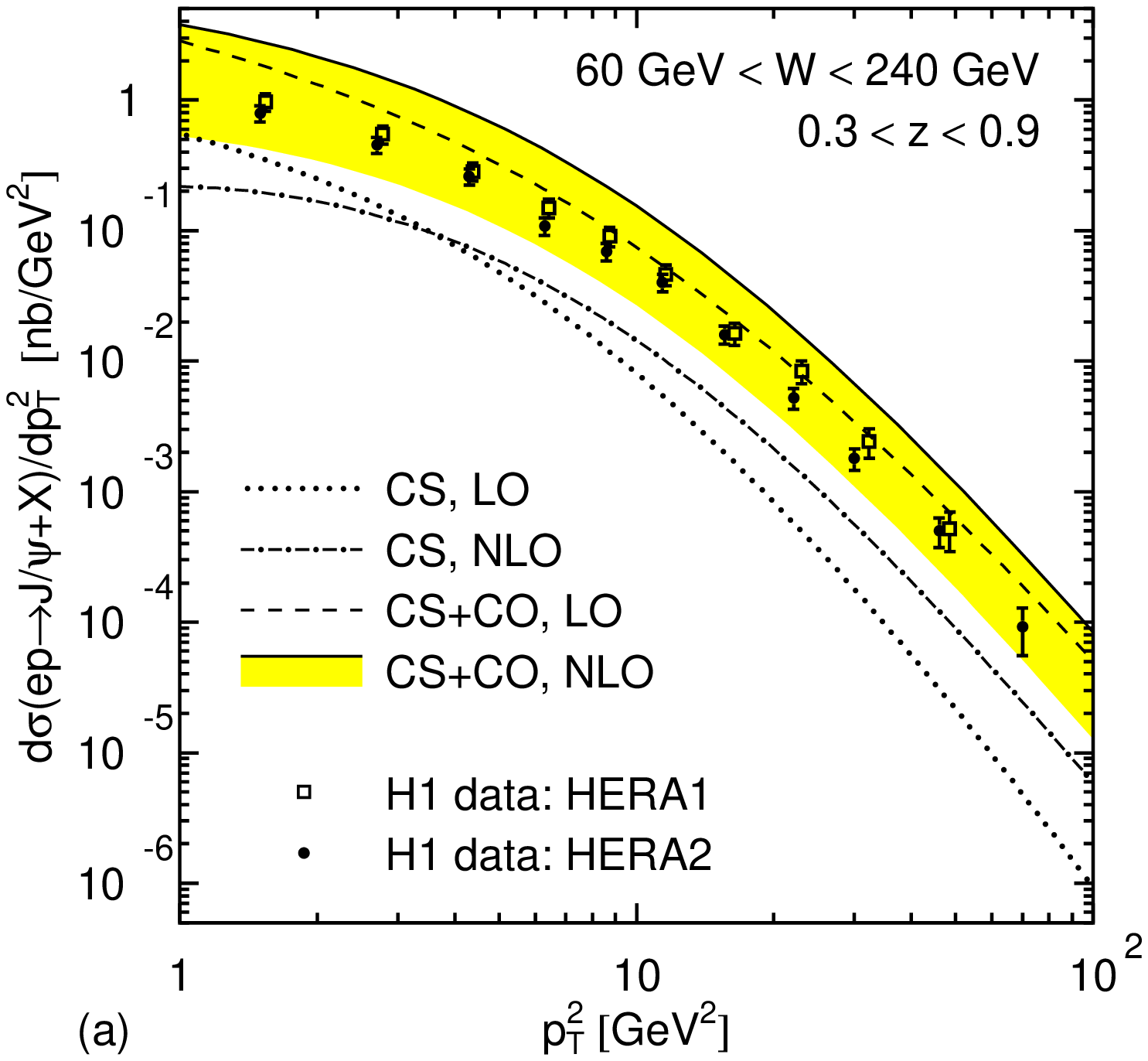}
&
\includegraphics[width=5cm]{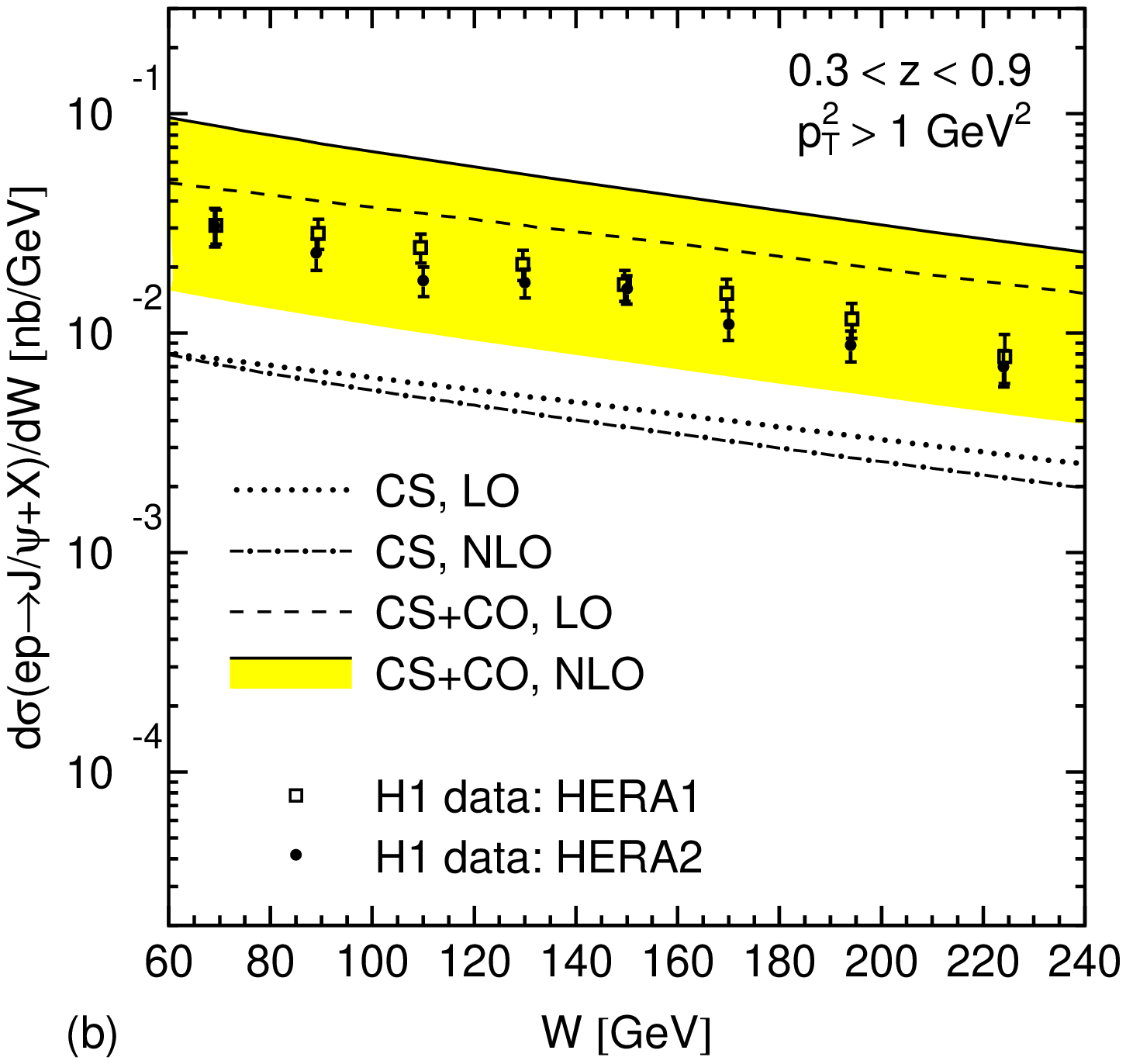}
&
\includegraphics[width=5cm]{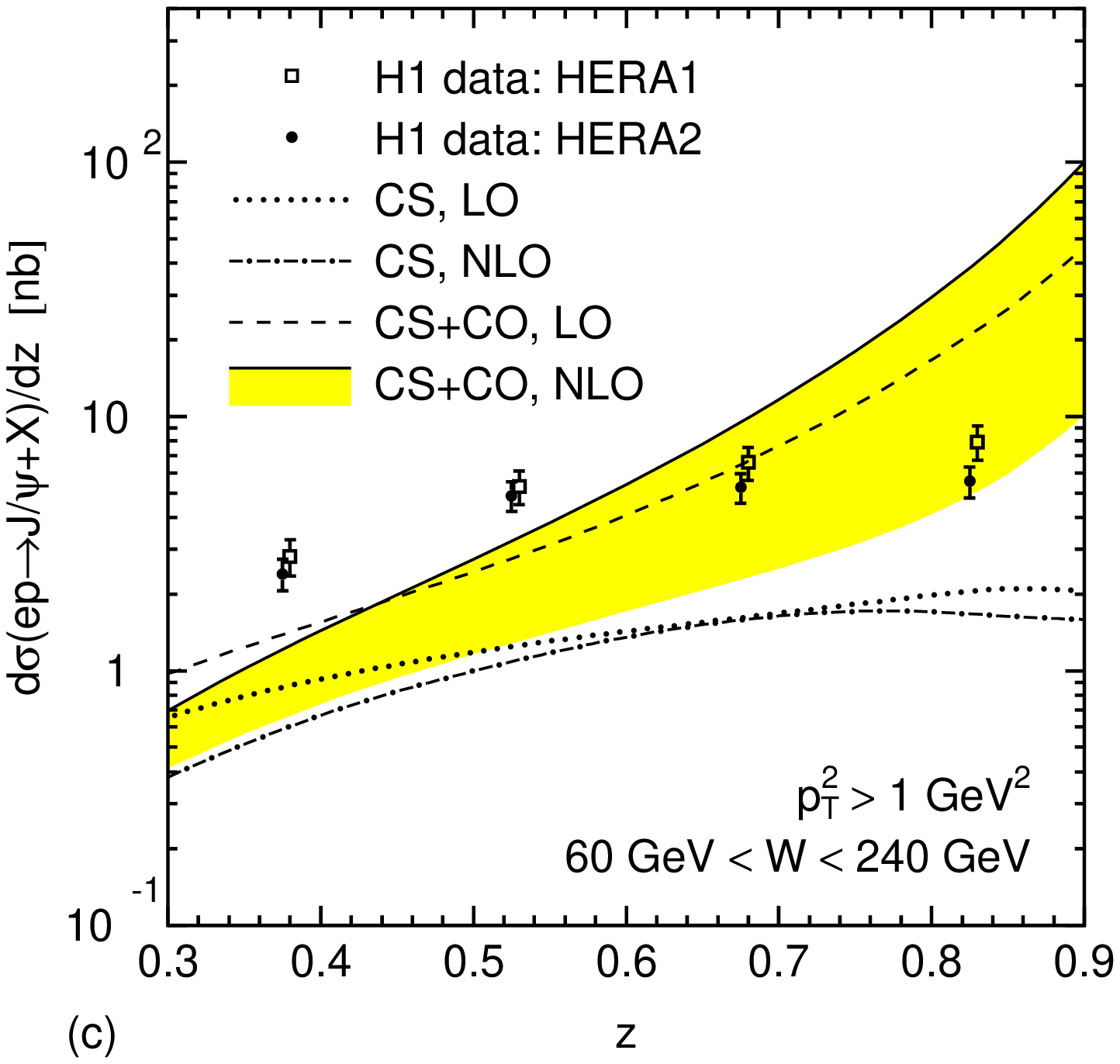}
\end{tabular}
\caption{\label{fig:results}
(a) $p_T^2$, (b) $W$, and (c) $z$ distributions of inclusive $J/\psi$
photoproduction at LO and NLO in the CSM and full NRQCD in comparison with H1
data \cite{Adloff:2002ex,H1.prelim}.
%\cite{Adloff:2002ex}.
The shaded (yellow) bands indicate the theoretical uncertainty due to the CO
LDMEs.}
\end{figure*}

We start our numerical analysis by estimating the theoretical uncertainties.
The dependences on the unphysical scales $\mu_r$ and $\mu_f$ are investigated
in full NRQCD at LO and NLO for the typical case of
$d\sigma(ep\to J/\psi+X)/dp_T^2$ at $p_T^2=20$~GeV$^2$ in
Fig.~\ref{fig:scales}.
Contrary to na\"\i ve expectations, the scale dependence is not reduced when
passing from LO to NLO.
Detailed investigation reveals that this behavior may be ascribed to the fact
that the new coefficient of $\alpha_s^3(\mu_r)$ is greatly dominated by the
part that does not carry logarithmic dependence on $\mu_r$ or $\mu_f$, mainly
arising from the gluon-induced ${}^1\!S_0^{[8]}$ and ${}^3\!P_0^{[8]}$
channels, while the complementary part still formally warrants renormalization
group invariance up to terms beyond NLO.
As for the dependence on $m_c$, a reduction of $m_c$ from
$m_{J/\psi}/2\approx1.55$~GeV to 1.4~GeV typically entails a rise in cross
section by about 50\%.
The freedom in sharing $M_r^{J/\psi}$ of Eq.~(\ref{eq:mr}) between
$\langle {\cal O}^{J/\psi}(^1\!S_0^{[8]}) \rangle$ and 
$(r/m_c^2)\langle {\cal O}^{J/\psi}(^3\!P_0^{[8]}) \rangle$ typically
creates an uncertainty of about 10\%.
The bulk of the theoretical uncertainty is actually due to the lack of
knowledge of the complete NLO corrections to the cross section of inclusive
$J/\psi$ hadroproduction, which is instrumental for a reliable NLO fit of the
CO LDMEs to the Tevatron data.
As explained above, these corrections are expected to be dominated by positive
and sizable contributions from real QCD bremsstrahlung
\cite{Campbell:2007ws,Gong:2008sn,Gong:2008ft,CanoColoma:1997rn}, leading to a
significant reduction of the CO LDMEs \cite{Kniehl:1998qy}.
At present, the theoretical uncertainty in inclusive $J/\psi$ photoproduction
from this source may be conservatively estimated by comparing the full NRQCD
evaluations using the LO and higher-order-improved LDME sets of
Ref.~\cite{Kniehl:1998qy}, with the understanding that the former is bound to
overshoot a future evaluation with a genuine NLO set.
This kind of uncertainty is indicated in the remaining figures by shaded
(yellow) bands, whose upper margins (solid lines) refer to the LO set.

The H1 measurements \cite{Adloff:2002ex,H1.prelim}
%\cite{Adloff:2002ex}
of the $p_T^2$, $W$, and $z$
distributions of inclusive $J/\psi$ photoproduction are compared with our new
NLO predictions in full NRQCD in Fig.~\ref{fig:results}(a)--(c), respectively. 
For comparison, also the default predictions at LO (dashed lines) as well as
those of the CSM at NLO (dot-dashed lines) and LO (dotted lines) are shown.
Notice that the experimental data are contaminated by the feed-down from
heavier charmonia, mainly due to $\psi^\prime\to J/\psi+X$, which yields an
estimated enhancement by about 15\% \cite{Kramer:1994zi}.
Furthermore, our predictions do not include resolved photoproduction, which
contributes appreciably only at $z\alt0.3$ \cite{Kniehl:1998qy}, and
diffractive production, which is confined to the quasi-elastic domain at
$z\approx1$ and $p_T\approx0$.
These contributions are efficiently suppressed by the cut $0.3<z<0.9$ in
Figs.~\ref{fig:results}(a) and (b), so that our comparisons are indeed
meaningful.
We observe that the NLO corrections enhance the NRQCD cross section, by up to
115\%, in the kinematic range considered, except for $z\alt0.45$, where they
are negative.
As may be seen from Fig.~\ref{fig:results}(c), the familiar growth of the LO
NRQCD prediction in the upper endpoint region, leading to a breakdown at
$z=1$, is further enhanced at NLO.
The solution to this problem clearly lies beyond the fixed-order treatment and
may be found in soft collinear effective theory \cite{Fleming:2006cd}.
The experimental data are nicely gathered in the central region of the error
bands, except for the two low-$z$ points in Fig.~\ref{fig:results}(c), which
overshoot the NLO NRQCD prediction.
However, this apparent disagreement is expected to fade away once the
NLO-corrected NRQCD contribution due to resolved photoproduction is included.
In fact, the above considerations concerning the large size of the NLO
corrections to hadroproduction directly carry over to resolved
photoproduction, which proceeds through the same partonic subprocesses.
On the other hand, the default CSM predictions significantly undershoot the
experimental data, by typically a factor of 4, which has already been observed
in Ref.~\cite{Artoisenet:2009xh}.
Except for $p_T^2\agt4$~GeV$^2$, the situation is even deteriorated by the
inclusion of the NLO corrections.

Despite the caveat concerning our limited knowledge of the CO LDMEs at NLO,
we conclude that the H1 data \cite{Adloff:2002ex,H1.prelim} show clear evidence
of the existence of CO processes in nature, as predicted by NRQCD, supporting
the conclusions previously reached for hadroproduction at the Tevatron
\cite{Cho:1995vh} and two-photon collisions at LEP2 \cite{Klasen:2001cu}.
%In order to further substantiate this argument, it is indispensable to
%determine the relevant CO LDMEs with NLO precision.
%Since the tightest constraints have so far come from the Tevatron and soon
%will from the CERN LHC, the most urgent next step is to complete the NLO
%analysis of inclusive $J/\psi$ hadroproduction in NRQCD, by treating also the
%$^3\!P_J^{[8]}$ channels at NLO.
In order to further substantiate this argument, it is indispensable to
complete the NLO analysis of inclusive $J/\psi$ hadroproduction in NRQCD, by
treating also the $^3\!P_J^{[8]}$ channels at NLO, so as to permit a genuine
NLO fit of the relevant CO LDMEs to Tevatron and CERN LHC data.
This goal is greatly facilitated by the technical advancement achieved in the
present analysis.

We thank L. Mihaila and J. Soto for useful discussions, and M. Steder for help
with the comparison to H1 data \cite{H1.prelim}.
This work was supported in part by BMBF Grant No.\ 05H09GUE, DFG Grant
No.\ KN~365/6--1, and HGF Grant No.\ HA~101.


\begin{thebibliography}{99}

\bibitem{Bodwin:1994jh}
  G.~T.~Bodwin, E.~Braaten, and G.~P.~Lepage,
  %``Rigorous QCD analysis of inclusive annihilation and production of heavy
  %quarkonium,''
  Phys.\ Rev.\  D {\bf 51}, 1125 (1995);
  {\bf 55}, 5853(E) (1997).
%  [arXiv:hep-ph/9407339].

\bibitem{Nayak:2005rt}
  G.~C.~Nayak, J.~W.~Qiu, and G.~Sterman,
  %``Fragmentation, NRQCD and NNLO factorization analysis in heavy  quarkonium
  %production,''
  Phys.\ Rev.\  D {\bf 72}, 114012 (2005);
%  [arXiv:hep-ph/0509021].
%\bibitem{Nayak:2006fm}
%  G.~C.~Nayak, J.~W.~Qiu and G.~Sterman,
  %``NRQCD Factorization and Velocity-dependence of NNLO Poles in Heavy
  %Quarkonium Production,''
%  Phys.\ Rev.\  D 
{\bf 74}, 074007 (2006).
%  [arXiv:hep-ph/0608066].

\bibitem{Cho:1995vh}
  P.~L.~Cho and A.~K.~Leibovich,
  %``Color octet quarkonia production,''
  Phys.\ Rev.\  D {\bf 53}, 150 (1996);
%  P.~L.~Cho and A.~K.~Leibovich,
  %``Color-octet quarkonia production II,''
  {\bf 53}, 6203 (1996).

\bibitem{Campbell:2007ws}
  J.~Campbell, F.~Maltoni, and F.~Tramontano,
  %``QCD corrections to J/psi and Upsilon production at hadron colliders,''
  Phys.\ Rev.\ Lett.\  {\bf 98}, 252002 (2007);
%  [arXiv:hep-ph/0703113].
%\bibitem{Artoisenet:2007xi}
  P.~Artoisenet, J.~P.~Lansberg, and F.~Maltoni,
  %``Hadroproduction of $J/\psi$ and $\upsilon$ in association with a
  %heavy-quark pair,''
  Phys.\ Lett.\  B {\bf 653}, 60 (2007);
%  [arXiv:hep-ph/0703129].
%\bibitem{Artoisenet:2008zza}
  P.~Artoisenet,
  %``QCD corrections to heavy quarkonium production,''
  AIP Conf.\ Proc.\  {\bf 1038}, 55 (2008).

\bibitem{Gong:2008sn}
  B.~Gong and J.-X.~Wang,
  %``Next-to-leading-order QCD corrections to $J/\psi$ polarization at Tevatron
  %and Large-Hadron-Collider energies,''
  Phys.\ Rev.\ Lett.\  {\bf 100}, 232001 (2008).
%  [arXiv:0802.3727 [hep-ph]].

\bibitem{Artoisenet:2009xh}
  P.~Artoisenet, J.~Campbell, F.~Maltoni, and F.~Tramontano,
  %``J/psi production at HERA,''
  Phys.\ Rev.\ Lett.\  {\bf 102}, 142001 (2009);
%  [arXiv:0901.4352 [hep-ph]];
%\bibitem{Chang:2009uj}
  C.-H.~Chang, R.~Li and J.-X.~Wang,
  %``J/\psi polarization in photo-production up-to the next-to-leading order of
  %QCD,''
  Phys.\ Rev.\  D {\bf 80}, 034020 (2009).
%  [arXiv:0901.4749 [hep-ph]].

\bibitem{Cacciari:1996dg}
  M.~Cacciari and M.~Kr\"amer,
  %``Color octet contributions to $J/\psi$ photoproduction,''
  Phys.\ Rev.\ Lett.\  {\bf 76}, 4128 (1996);
%  [arXiv:hep-ph/9601276].
%\cite{Ko:1996xw}
%\bibitem{Ko:1996xw}
  P.~Ko, J.~Lee, and H.~S.~Song,
  %``Color octet mechanism in $\gamma$ + $p \to J/\psi$ + x,''
  Phys.\ Rev.\  D {\bf 54}, 4312 (1996);
  {\bf 60}, 119902(E) (1999).
%  [arXiv:hep-ph/9602223].

\bibitem{Kniehl:2001tk}
  B.~A.~Kniehl and L.~Zwirner,
  %``$J/\psi$ inclusive production in $e p$ deep inelastic scattering at DESY
  %HERA,''
  Nucl.\ Phys.\  B {\bf 621}, 337 (2002);
%  [arXiv:hep-ph/0112199].

\bibitem{Klasen:2001cu}
  M.~Klasen, B.~A.~Kniehl, L.~N.~Mihaila, and M.~Steinhauser,
  %``Evidence for color octet mechanism from CERN LEP-2 $\gamma \gamma \to
  %J/\psi$ + $X$ data,''
  Phys.\ Rev.\ Lett.\  {\bf 89}, 032001 (2002).
%  [arXiv:hep-ph/0112259].

%\bibitem{Zhang:2006ay}
%  Y.-J.~Zhang and K.-T.~Chao,
  %``Double charm production e+ e- --> J/psi + c anti-c at B factories with
  %next-to-leading order QCD correction,''
%  Phys.\ Rev.\ Lett.\  {\bf 98}, 092003 (2007).
%  [arXiv:hep-ph/0611086].

%\bibitem{Ma:2008gq}
%  Y.-Q.~Ma, Y.-J.~Zhang, and K.-T.~Chao,
  %``QCD correction to $\bm{e^+ e^- \to J/\psi g g}$ at B Factories,''
%  Phys.\ Rev.\ Lett.\  {\bf 102}, 162002 (2009);
%  [arXiv:0812.5106 [hep-ph]].
%\bibitem{Gong:2009kp}
%  B.~Gong and J.~X.~Wang,
  %``Next-to-Leading-Order QCD Corrections to e^ + e^- -> J/\psi+gg at the B
  %Factories,''
%  Phys.\ Rev.\ Lett.\  
%{\it ibid.}\ {\bf 102}, 162003 (2009).
%  [arXiv:0901.0117 [hep-ph]].

\bibitem{Braaten:1999qk}
  E.~Braaten, B.~A.~Kniehl, and J.~Lee,
  %``Polarization of prompt $J/\psi$ at the Tevatron,''
  Phys.\ Rev.\  D {\bf 62}, 094005 (2000);
%  [arXiv:hep-ph/9911436].
%\bibitem{Kniehl:2000nn}
  B.~A.~Kniehl and J.~Lee,
  %``Polarized $J/\psi$ from chi(cJ) and $\psi^\prime$ decays at the Tevatron,''
%  Phys.\ Rev.\  D
 {\it ibid.}\ {\bf 62}, 114027 (2000).
%  [arXiv:hep-ph/0007292].

\bibitem{Gong:2008ft} %NLO hadroprod in CSM w/ S-wave CO xs + pol
  B.~Gong, X.~Q.~Li, and J.-X.~Wang,
  %``QCD corrections to $J/\psi$ production via color octet states at Tevatron
  %and LHC,''
  Phys.\ Lett.\  B {\bf 673}, 197 (2009).
%  [arXiv:0805.4751 [hep-ph]].

\bibitem{Kramer:1994zi}
  M.~Kr\"amer, J.~Zunft, J.~Steegborn, and P.~M.~Zerwas,
  %``Inelastic J / psi photoproduction,''
  Phys.\ Lett.\  B {\bf 348}, 657 (1995);
%  [arXiv:hep-ph/9411372];
%\bibitem{Kramer:1995nb}
  M.~Kr\"amer,
  %``QCD Corrections to Inelastic $J/\psi$ Photoproduction,''
  Nucl.\ Phys.\  B {\bf 459}, 3 (1996).
%  [arXiv:hep-ph/9508409].

\bibitem{Petrelli:1997ge}
  A.~Petrelli, M.~Cacciari, M.~Greco, F.~Maltoni, and M.~L.~Mangano,
  %``NLO production and decay of quarkonium,''
  Nucl.\ Phys.\  B {\bf 514}, 245 (1998)
%  [arXiv:hep-ph/9707223].

\bibitem{Klasen:2004tz}
  M.~Klasen, B.~A.~Kniehl, L.~N.~Mihaila, and M.~Steinhauser,
  %``$J/\psi$ plus jet associated production in two-photon collisions at
  %next-to-leading order,''
  Nucl.\ Phys.\  B {\bf 713}, 487 (2005);
%  [arXiv:hep-ph/0407014].
%\bibitem{Klasen:2004az}
%  M.~Klasen, B.~A.~Kniehl, L.~N.~Mihaila and M.~Steinhauser,
  %``$J/\psi$ plus prompt-photon associated production in two-photon collisions
  %at next-to-leading order,''
  Phys.\ Rev.\  D {\bf 71}, 014016 (2005).
%  [arXiv:hep-ph/0408280].

%\bibitem{von Weizsacker:1934sx}
%  E.~J.~Williams, Proc.\ R.\ Soc.\ London\ (A) {\bf 139}, 163 (1933);
%  C.~F.~v.~Weizs\"acker,
%  %``Radiation emitted in collisions of very fast electrons,''
%  Z.\ Phys.\  {\bf 88}, 612 (1934).

%\bibitem{Maltoni:2002qb}
%  F.~Maltoni and T.~Stelzer,
%  %``MadEvent: Automatic event generation with MadGraph,''
%  JHEP {\bf 0302}, 027 (2003).
%%  [arXiv:hep-ph/0208156].

\bibitem{Artoisenet:2007qm}
  P.~Artoisenet, F.~Maltoni, and T.~Stelzer,
  %``Automatic generation of quarkonium amplitudes in NRQCD,''
  JHEP {\bf 0802}, 102 (2008).
%  [arXiv:0712.2770 [hep-ph]].
  %%CITATION = JHEPA,0802,102;%%

%\bibitem{Harris:2001sx}
%  B.~W.~Harris and J.~F.~Owens,
  %``The two cutoff phase space slicing method,''
%  Phys.\ Rev.\  D {\bf 65}, 094032 (2002).
%  [arXiv:hep-ph/0102128].

\bibitem{Amsler:2008zzb}
  Particle Data Group, C.~Amsler {\it et al.},
  %``Review of particle physics,''
  Phys.\ Lett.\  B {\bf 667}, 1 (2008).

\bibitem{Pumplin:2002vw}
CTEQ Colaboration, J.~Pumplin {\it et al.},
%, D.~R.~Stump, J.~Huston, H.~L.~Lai, P.~M.~Nadolsky and W.~K.~Tung,
  %``New generation of parton distributions with uncertainties from global QCD
  %analysis,''
  JHEP {\bf 0207}, 012 (2002).
%  [arXiv:hep-ph/0201195].

\bibitem{Kniehl:1996we}
  B.~A.~Kniehl, G.~Kramer, and M.~Spira,
  %``Large p(T) photoproduction of D*+- mesons in e p collisions,''
  Z.\ Phys.\  C {\bf 76}, 689 (1997).
%  [arXiv:hep-ph/9610267].
  
\bibitem{Adloff:2002ex}
  H1 Collaboration, C.~Adloff {\it et al.},
  %``Inelastic photoproduction of $J/\psi$ mesons at HERA,''
  Eur.\ Phys.\ J.\  C {\bf 25}, 25 (2002).
%  [arXiv:hep-ex/0205064].

\bibitem{H1.prelim}
%  M. Steder, on behalf of the H1 Collaboration, Report No.\ H1prelim--07--172.
F.D. Aaron {\it et al.}\ (H1 Collaboration), DESY~09--225,
arXiv:1002.0234 [hep-ex].

\bibitem{Kniehl:1998qy}
  B.~A.~Kniehl and G.~Kramer,
  %``TEVATRON - HERA color - octet charmonium anomaly versus higher order QCD
  %effects,''
  Eur.\ Phys.\ J.\  C {\bf 6}, 493 (1999).
%  [arXiv:hep-ph/9803256].

\bibitem{CanoColoma:1997rn}
  B.~Cano-Coloma and M.~A.~Sanchis-Lozano,
  %``Charmonia production in hadron colliders and the extraction of color octet
  %matrix elements,''
  Nucl.\ Phys.\  B {\bf 508}, 753 (1997).
%  [arXiv:hep-ph/9706270].

\bibitem{Kniehl:2006qq}
  B.~A.~Kniehl and C.~P.~Palisoc,
  %``Prompt $J/\psi$ plus photon associated electroproduction at DESY HERA,''
  Eur.\ Phys.\ J.\  C {\bf 48}, 451 (2006).
%  [arXiv:hep-ph/0608245].
    

\bibitem{Fleming:2006cd}
  S.~Fleming, A.~K.~Leibovich, and T.~Mehen,
  %``Resummation of Large Endpoint Corrections to Color-Octet $J/\psi$
  %Photoproduction,''
  Phys.\ Rev.\  D {\bf 74}, 114004 (2006).
%  [arXiv:hep-ph/0607121].

\end{thebibliography}
\end{document}